\documentclass[aps,prl,showpacs,amsmath,amssymb,twocolumn,10pt]{revtex4-1} 
\usepackage{amsthm}
\usepackage{dsfont}
\usepackage{psfrag}
\usepackage[dvips]{graphicx}

\newtheorem{proposition}{Proposition}

\newcommand\RR{{\mathds{R}}}
\newcommand\CC{{\mathds{C}}}

\newcommand\ee{{\mathrm{e}}}
\newcommand\ii{{\mathrm{i}}}

\DeclareMathOperator{\Tr}{Tr}

\begin{document} 

\title{Diverging equilibration times in long-range quantum spin models} 

\author{Michael Kastner} 
\email{kastner@sun.ac.za} 
\affiliation{National Institute for Theoretical Physics (NITheP), Stellenbosch 7600, South Africa} 
\affiliation{Institute of Theoretical Physics,  University of Stellenbosch, Stellenbosch 7600, South Africa}

\date{\today}
 
\begin{abstract}
The approach to equilibrium is studied for long-range quantum Ising models where the interaction strength decays like $r^{-\alpha}$ at large distances $r$ with an exponent $\alpha$ not exceeding the lattice dimension. For a large class of observables and initial states, the time evolution of expectation values can be calculated. We prove analytically that, at a given instant of time $t$ and for sufficiently large system size $N$, the expectation value of some observable $\langle A\rangle(t)$ will practically be unchanged from its initial value $\langle A\rangle(0)$. This finding implies that, for large enough $N$, equilibration effectively occurs on a time scale beyond the experimentally accessible one and will not be observed in practice.
\end{abstract}


\maketitle 

Equilibration is one of the central concepts of thermodynamics, but our understanding of the underlying microscopic processes is still far from complete. Studies of the approach to equilibrium can be traced back to Boltzmann's work \cite{Boltzmann72} in the early days of statistical mechanics, and they are also closely related to the microscopic foundations of the second law of thermodynamics. Modern facets of this topic, attracting much attention in the recent research literature, are relaxation after a quantum quench \cite{Cramer_etal08,*Polkovnikov_etal}, thermalization close to integrability \cite{Kinoshita_etal06,*Rigol09,*Relano10}, or typicality as a foundation of quantum statistical mechanics \cite{Reimann08,*Linden_etal09,*Goldstein_etal10}, to name but a few.

Although experience shows that equilibration takes place in the vast majority of situations, exceptions have always attracted particular interest. A famous example is the numerical investigation of the time evolution of a chain of nonlinearly coupled oscillators by Fermi, Pasta, and Ulam \cite{FPU55} where recurrent behavior was observed, but no sign of equilibration. Another particularly interesting case, going under the name of quasistationary states, has triggered intense activity, reviewed in \cite{CamDauxRuf09}, in the field of long-range interacting systems. {\em Long-range}\/ refers here to interactions decaying at large distances $r$ as a power law $r^{-\alpha}$ with a positive exponent $\alpha$ not exceeding the spatial dimension $d$ of the system. Gravitating masses or Coulomb forces in the absence of screening are important examples of long-range interactions. The term {\em quasistationary}\/ is used to describe metastable states whose lifetime diverges with increasing system size $N$. The physical importance of quasistationary behavior should be obvious: For a sufficiently large system, the transition from a quasistationary state to equilibrium takes place on a time scale that is larger than the experimentally accessible observation time. Hence, equilibrium properties will not be observed, and instead the statistical properties of the quasistationary regime are of interest.

Most of the studies of quasistationary states have focused on the Hamiltonian Mean-Field model \cite{AnRu95}, a toy model consisting of classical $XY$ spins (or plane rotators), each coupled to every other at equal strength. The typical scenario observed is that, for a suitable class of initial conditions, the total magnetization of this spin model rapidly relaxes to some quasistationary value different from its equilibrium value, remains there for a very long time, and finally decays to equilibrium (see Fig.\ 33 in \cite{CamDauxRuf09}). More recently, quasistationary states have been observed in gravitational sheet models \cite{JoyceWorrakitpoonpon10} and their existence has been argued to be generic for a large class of classical long-range systems \cite{Gabrielli_etal10}. Virtually all finite-system results in the field have been obtained by numerical techniques, supplemented by analytical calculations in the $N\to\infty$ Vlasov continuum limit \cite{CamDauxRuf09}.

In this Letter, analytic results on equilibration in long-range quantum spin models are reported. Surprisingly, the class of models studied permits exact analytic results for both, finite and infinite systems, even beyond what has been achieved in the classical case. In principle, the time evolution is found to be recurrent for any finite system size $N$, and no equilibration occurs in a strict sense. But since the recurrence times diverge exponentially with $N$, they will be irrelevant for large enough systems and effective equilibration is observed. Similar to the quasistationarity in classical systems, we find that the system remains close to its initial state for extremely long times and effective equilibration times diverge with increasing system size. This is, to the best of the author's knowledge, the first observation of this kind of behavior in quantum systems, and it prepares the ground for further studies, including the above-mentioned quantum quenches or inquiries into the foundations of statistical physics of long-range systems. The results complement the peculiarities of long-range quantum spin systems in equilibrium reported in \cite{Kastner10}. 

Groundbreaking advances have been witnessed in recent years regarding the experimental realization of spin models, where ultracold gases in optical lattices have been used to engineer spin systems whose parameters can be tuned. Long-range interactions between the constituents can be created by either trapping dipolar molecules \cite{Micheli_etal06}, or by inducing a nonpermanent dipole moment when shining appropriately tuned laser light onto atoms \cite{ODell_etal00}. Future experiments of this kind might permit the experimental observation of diverging equilibration times in long-range quantum spin systems.

{\em The model.---}The main result of this Letter can be proved for a large class of (generalized) Ising models with long-range pair interactions, defined on various types of lattices and for arbitrary lattice dimensions. To keep the presentation simple, we discuss only the one-dimensional case explicitly. Higher-dimensional lattices and other generalizations are discussed below.

Consider $N$ identical spin-$1/2$ particles, attached to the sites $\{1,\dotsc,N\}$ of a finite one-dimensional lattice with periodic boundary conditions. The corresponding quantum dynamics takes place on the Hilbert space
\begin{equation}
\mathcal{H}=\bigotimes_{i=1}^N \CC_i^2,
\end{equation}  
where the $\CC_i^2$ are identical replicas of the two-dimensional Hil\-bert space of a single spin-$1/2$ particle. The time evolution on $\mathcal{H}$ is generated by the Hamiltonian
\begin{equation}\label{eq:Hamiltonian}
H_N=\mathcal{N}_N\sum_{i=1}^N\sum_{j=1}^{N/2} \epsilon(j)\sigma_i^z \sigma_{i+j}^z - h\sum_{i=1}^N \sigma_i^z,
\end{equation}
with $\sigma^z$ denoting the $z$ component of the Pauli spin operator and $h\in\RR$ an external magnetic field in the $z$ direction. The index $i+j$ is to be considered modulo $N$ to account for the periodic boundary conditions. The $\epsilon(j)$ are pair coupling constants depending on the distance $j$ of two spin operators on the lattice, and we assume
\begin{equation}
\lim_{j\to\infty}\epsilon(j)=0.
\end{equation}
A typical example we have in mind is the Dyson model \cite{Dyson69a} with algebraically decaying couplings, $\epsilon(j)=j^{-\alpha}$, but, in contrast to Dyson's work, exponents $\alpha$ smaller than 1 are also allowed. More general classes of interactions will be discussed below. For exponents $\alpha>1$, the interaction is absolutely summable,
\begin{equation}\label{eq:l1}
\sum_{j=1}^\infty|\epsilon(j)|<\infty,
\end{equation}
and this case has already been treated in \cite{Emch66,Radin70,Wreszinski10}. Here we are interested in what is sometimes called {\em strong}\/ long-range interactions with $0<\alpha<1$ where the sum in \eqref{eq:l1} diverges. In this case, the double sum in \eqref{eq:Hamiltonian} gives rise to an infinite energy per spin $\langle H_N\rangle/N$ in the thermodynamic limit $N\to\infty$. To render this limit finite, the normalization
\begin{equation}
\mathcal{N}_N=\Biggl(2\sum_{j=1}^{N/2} \epsilon(j)\Biggr)^{-1}
\end{equation}
has been introduced in \eqref{eq:Hamiltonian}. This normalization is a generalization of the so-called Kac prescription commonly used when studying long-range interacting systems. For large system sizes $N$, its asymptotic behavior is of the form
\begin{equation}\label{eq:Nasymptotic}
2\mathcal{N}_N\sim
\begin{cases}
(1-\alpha)2^{1-\alpha}N^{\alpha-1} &\text{for $0\leqslant\alpha<1$},\\
1/\ln N &\text{for $\alpha=1$},\\
1/\zeta(\alpha) &\text{for $\alpha>1$},
\end{cases}
\end{equation}
where $\zeta$ denotes the Riemann zeta function. Equation \eqref{eq:Nasymptotic} implies that $\mathcal{N}_N$ vanishes in the thermodynamic limit for exponents $\alpha\leqslant1$.

In most studies of the (generalized) Ising model \eqref{eq:Hamiltonian}, observables related to the $z$ components $\sigma_i^z$ are considered, like the magnetization per spin in the $z$ direction, $\sum_{i=1}^N \sigma_i^z/N$. Since observables of this type commute with $H_N$, their expectation values are conserved quantities and equilibration cannot be observed. Here we want to consider the time-dependence of the expectation value of observables which are linear combinations of the $x$ components $\sigma_i^x$ of the spin operators,
\begin{equation}\label{eq:A}
A(a_1,\dotsc,a_N)=\sum_{i=1}^N a_i \sigma_i^x
\end{equation}
with real coefficients $a_i$. None of these observables commutes with the Hamiltonian \eqref{eq:Hamiltonian}. The set containing all $A$ is, as pointed out in \cite{Emch66}, maximal Abelian, meaning that the information obtained by measuring simultaneously all $A$ cannot be improved by any compatible measurement performed at the same instant. Our aim is to study the time evolution of observables $A$, starting from an initial state (i.e., density operator) $\rho(0)$ which is diagonal in the $\sigma_i^x$ tensor product eigenbasis of $\mathcal{H}$.

{\em Related models and earlier results.---}The above model is inspired by work of Emch \cite{Emch66} and Radin \cite{Radin70}. The crucial difference of our model, compared to the work of Emch and Radin, is the presence of strong long-range interactions and of the $N$-dependent normalization factor $\mathcal{N}_N$ in the Hamiltonian \eqref{eq:Hamiltonian}, and we will see that these ingredients lead to dramatically different behavior.

The model introduced in \cite{Emch66}, including the choices of observables and initial states, is inspired by induction decay experiments probing the pulsed magnetic resonance of nuclei in a CaF$_2$ crystal \cite{LoweNorberg57}. In these experiments, the decay of the $x$ component of the total magnetization $A(1,\dotsc,1)= \sum_{i=1}^N \sigma_i^x$ was measured, and it was found to be superimposed by oscillations (see Fig.\ 1 of \cite{LoweNorberg57}). Despite its simplifying assumptions, the model captures well both the decay and the beating observed experimentally. Moreover, the Emch-Radin model and its random generalizations \cite{Wreszinski10} have proved useful as paradigmatic models for which the approach to equilibrium in quantum systems can be studied analytically.

\begin{figure}\center
\includegraphics[width=0.95\linewidth]{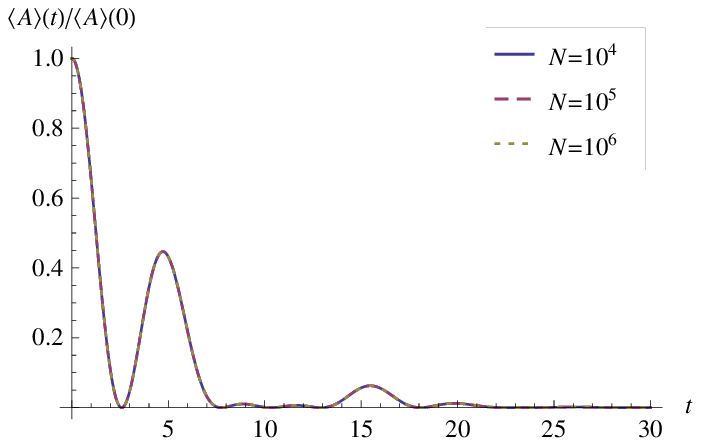}\\
\includegraphics[width=0.95\linewidth]{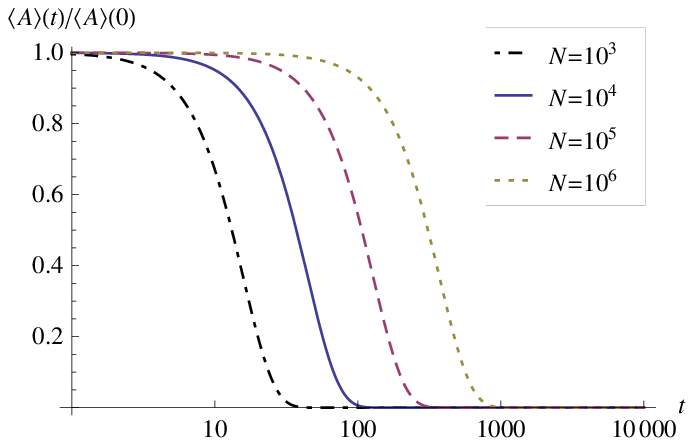}
\caption{\label{fig:decay}
Time evolution of the expectation value \eqref{eq:A_of_t} of an observable $A$ for magnetic field $h=0$ and various system sizes $N$. Top panel: For $\alpha=2$, an apparent decay is observed, superimposed by oscillations. The $N$ dependence of the time evolution is so weak that the curves for the various system sizes cannot be discerned in the plot. The behavior for other $\alpha>1$ is qualitatively similar. Bottom panel: For $\alpha=1/2$, the expectation value again appears to be decaying, but on a time scale that depends strongly on the system size $N$ (note the logarithmic scale). Similar behavior is observed for other values of $\alpha$ between zero and one.
}
\end{figure}

{\em Time evolution.---}We study the time evolution, generated by $H_N$, of the expectation value of an observable $A$ of the form \eqref{eq:A} with respect to the initial state $\rho(0)$,
\begin{equation}
\langle A\rangle(t)=\Tr\left[\ee^{-\ii H_N t} A \ee^{\ii H_N t}\rho(0)\right]
\end{equation}
(in units where $\hbar=1$). Performing the trace in the $\sigma_i^x$ eigenbasis of $\mathcal{H}$, the diagonal form of $\rho(0)$ implies that only the diagonal elements of the operator $\ee^{-\ii H_N t} A \ee^{\ii H_N t}$ are required, and it is this crucial ingredient which allows us to obtain, similar to the calculation in \cite{Emch66}, the exact result
\begin{equation}\label{eq:A_of_t}
\langle A\rangle(t)=\langle A\rangle(0)\cos(2ht)\prod_{j=1}^N \cos^2[2\mathcal{N}_N\epsilon(j)t].
\end{equation}
In comparison with the original Emch-Radin model, the important difference here is the explicit $N$ dependence of the argument of the cosine through the normalization $\mathcal{N}_N$. Regarding the approach to equilibrium, the Larmor precession $\cos(2ht)$ is not relevant and we set $h=0$ in the following. The behavior of \eqref{eq:A_of_t} is plotted for exponents $\alpha=2$ and $\alpha=1/2$ and various system sizes $N$ in Fig.\ \ref{fig:decay}, and in all cases the expectation value of $A$ appears to be decaying in time to the microcanonical ensemble average $\langle A\rangle_\text{mic}=0$.%
However, this decay is only apparent, as we can read off from \eqref{eq:A_of_t} that $\langle A\rangle(t)$ is an almost periodic function in time for all finite $N$, and Poincar\'e recurrences will therefore occur on much longer time scales than shown. To possibly observe true equilibration, we have to invoke, as often in statistical physics, the idealizing concept of the thermodynamic limit. In this limit, recurrence times may diverge (typically exponentially) and an approach to equilibrium may take place. An analysis of the infinite system dynamics is most rigorously done in a $\ast$-algebraic language \cite{Radin70,Wreszinski10}, but it essentially boils down to discussing the large-$N$ limit of the product in \eqref{eq:A_of_t}. Similar to Lemma 4 of \cite{Radin70}, we obtain the upper bound
\begin{equation}\label{eq:A_bound}
\lim_{N\to\infty}\left|\langle A\rangle(t)\right|\leqslant\left|\langle A\rangle(0)\right|
\exp\bigl\{-c\mathcal{N}_\infty t^2\bigr\}
\end{equation}
where $c$ is a positive constant. For $\alpha>1$, the large-$N$ limit of the normalization $\mathcal{N}_\infty$ in \eqref{eq:Nasymptotic} is strictly positive, proving a stretched exponential approach of $\langle A\rangle$ to its equilibrium value. For the case $0\leqslant\alpha<1$ of strong long-range interactions we are particularly interested in, $\mathcal{N}_\infty$ is zero. The bound in \eqref{eq:A_bound}, therefore, gives the positive constant $\left|\langle A\rangle(0)\right|$ and fails to provide any indication on whether equilibrium is approached or not. Up to this point the analysis has been very much along the lines of earlier work reported in the literature. The main novel result of the present Letter is to complement the upper bound \eqref{eq:A_bound} by a lower bound on $\langle A\rangle$ proving and characterizing the divergent equilibration time for $0\leqslant\alpha<1$ in the large-$N$ limit.
\begin{proposition}\label{prop1}
Consider the Ising-type Hamiltonian $H_N$ defined in \eqref{eq:Hamiltonian} with power law decaying interactions $\epsilon(j)=j^{-\alpha}$ with $0\leqslant\alpha<1$. Consider further an observable $A$ of the type \eqref{eq:A} and an initial state $\rho(0)$ being diagonal in the $\sigma_i^x$ tensor product eigenbasis of the underlying Hilbert space $\mathcal{H}$. Then for the expectation value $\langle A\rangle(t)$ of $A$ with respect to $\rho(t)$, the following holds true: For any fixed time $\tau$ and some small $\delta>0$, there is a finite $N_0(\tau)$ such that
\begin{equation}
\left|\langle A\rangle(t)-\langle A\rangle(0)\right|<\delta\qquad\text{$\forall t<\tau$, $N>N_0(\tau)$}.
\end{equation}
\end{proposition}
Interpreting this result in terms of an experiment, we can think of an experimental resolution $\delta$ for the measurement of $A$, and some duration $\tau$ of the experiment. Then the above proposition states that, within the experimental resolution and for a large enough system, no deviation of $\langle A\rangle(t)$ from its initial value can be observed for times $t\leqslant\tau$. In the above sense, $\langle A\rangle(t)$ converges in the thermodynamic limit to the constant $\langle A\rangle(0)$ which, in general, is different from the microcanonical ensemble average $\langle A\rangle_\text{mic}=0$.

Proposition \ref{prop1} can be proved by constructing, for large enough $N$, a lower bound on the product in \eqref{eq:A_of_t} by means of an integral approximation. The result can be expressed in terms of hypergeometric functions and, by virtue of the asymptotic properties of these functions, the bound can be pushed arbitrarily close to 1 by increasing the system size $N$.

Comparing the scaling of the equilibration times $\tau_0$ with system size $N$, 
a striking difference between classical and quantum mechanical quasistationary behavior is revealed: From the data plotted in Fig.\ \ref{fig:decay} (bottom) and consistent with the bound \eqref{eq:A_bound}, one finds $\tau_0\propto N^\gamma$ with $\gamma\approx0.5$, whereas in the classical case $\gamma\approx1.7$ has been extracted from numerical data and $\gamma=1$ from a Vlasov approach \cite{BuylMukamelRuffo}.

{\em Generalizations.---}To keep the presentation simple, the above exposition had been restricted to a one-dimensional lattice and power law interactions $\epsilon(j)=j^{-\alpha}$, but several generalizations are straightforward. In fact, only the large-$j$ asymptotic behavior of $\epsilon(j)$ is relevant for the proof of Proposition \ref{prop1}, and the result holds for any $|\epsilon(j)|$ decaying asymptotically as $Cj^{-\alpha}$ with some positive constant $C$ and $0<\alpha<1$. For lattice dimensions $d$, the results remain true for exponents $0<\alpha<d$, again with only marginal modifications of the proof.

Instead of observables $A$ of the type \eqref{eq:A} we considered, linear combinations of $y$ components of Pauli spin operators would not have altered any of the conclusions. Linear combinations of $z$ components, however, commute with the Hamiltonian \eqref{eq:Hamiltonian} and are therefore conserved quantities which do not equilibrate. The results in \cite{Radin70} indicate that even further generalizations 
should be feasible. 

{\em Discussion and outlook.---}We have analytically studied the time evolution of long-range quantum spin models with Ising-type interactions \eqref{eq:Hamiltonian} on lattices of arbitrary dimension and for a certain class of initial states. For finite system sizes $N$, almost periodic behavior is observed with recurrence times that increase exponentially with $N$. In a rigorous sense, equilibration can therefore occur only in the thermodynamic limit of infinite lattice size. Whether this indeed happens depends on the asymptotic behavior of the interaction strength $\epsilon(j)$ for large distances $j$ on the lattice. For interactions decaying faster than $Cj^{-\alpha}$ with some constant $C$ and exponent $\alpha>1$, equilibration takes place as expected and the expectation value $\langle A\rangle(t)$ approaches the microcanonical ensemble average $\langle A\rangle_\text{mic}=0$ exponentially for large $t$.

For exponents $\alpha$ between zero and one, the interaction strength $\epsilon(j)$ is not absolutely summable and an $N$-dependent normalization $\mathcal{N}_N$ is needed in \eqref{eq:Hamiltonian} to render the energy per spin finite. In this case, our Proposition \ref{prop1} asserts that, at a given instant of time and for large enough $N$, $\langle A\rangle(t)$ will practically be unchanged from its initial value $\langle A\rangle(0)$. In other words, for large enough systems, equilibration occurs on a time scale beyond the experimentally accessible one and will not be observed in practice. Despite the superficial similarities, such a behavior is notably different from the failure of thermalization reported for near-integrable systems \cite{Kinoshita_etal06,*Rigol09,*Relano10}. Our results extend the concept of quasistationary states, previously observed and extensively studied in classical systems, into the realm of quantum mechanics, while at the same time providing the first rigorous, analytic proof of quasistationary behavior from the microscopic time-evolution equations.

It is tempting to speculate that quasistationary behavior, i.e., equilibration times that diverge for large system sizes $N$, might show up under much more general conditions, or even generically for long-range interacting quantum systems. An extension of our results to general anisotropic Heisenberg models might be a promising future project and, although an exact solution for the time evolution of $\langle A\rangle$ seems to be out of reach, the derivation of suitable bounds on $\langle A\rangle(t)$ might be feasible. 
Finally, the simultaneous presence of short- and long-range interactions might lead to equilibration taking place on two different time scales, possibly leading to more complex behavior. The peculiarities of diverging relaxation times must be expected to be reflected also in many applied aspects of quantum spin systems, including the study of quantum quenches \cite{Cramer_etal08,*Polkovnikov_etal}, decoherence, and many others.

Financial support by the {\em Incentive Funding for Rated Researchers}\/ programme of the National Research Foundation of South Africa is gratefully acknowledged.

\bibliography{QQSS.bib}

\begin{thebibliography}{23}%
\makeatletter
\providecommand \@ifxundefined [1]{%
 \@ifx{#1\undefined}
}%
\providecommand \@ifnum [1]{%
 \ifnum #1\expandafter \@firstoftwo
 \else \expandafter \@secondoftwo
 \fi
}%
\providecommand \@ifx [1]{%
 \ifx #1\expandafter \@firstoftwo
 \else \expandafter \@secondoftwo
 \fi
}%
\providecommand \natexlab [1]{#1}%
\providecommand \enquote  [1]{``#1''}%
\providecommand \bibnamefont  [1]{#1}%
\providecommand \bibfnamefont [1]{#1}%
\providecommand \citenamefont [1]{#1}%
\providecommand \href@noop [0]{\@secondoftwo}%
\providecommand \href [0]{\begingroup \@sanitize@url \@href}%
\providecommand \@href[1]{\@@startlink{#1}\@@href}%
\providecommand \@@href[1]{\endgroup#1\@@endlink}%
\providecommand \@sanitize@url [0]{\catcode `\\12\catcode `\$12\catcode
  `\&12\catcode `\#12\catcode `\^12\catcode `\_12\catcode `\%12\relax}%
\providecommand \@@startlink[1]{}%
\providecommand \@@endlink[0]{}%
\providecommand \url  [0]{\begingroup\@sanitize@url \@url }%
\providecommand \@url [1]{\endgroup\@href {#1}{\urlprefix }}%
\providecommand \urlprefix  [0]{URL }%
\providecommand \Eprint [0]{\href }%
\@ifxundefined \urlstyle {%
  \providecommand \doi  [0]{\begingroup \@sanitize@url \@doi}%
  \providecommand \@doi [1]{\endgroup \@@startlink {\doibase
  #1}doi:\discretionary {}{}{}#1\@@endlink }%
}{%
  \providecommand \doi  [0]{doi:\discretionary{}{}{}\begingroup
  \urlstyle{rm}\Url }%
}%
\providecommand \doibase [0]{http://dx.doi.org/}%
\providecommand \Doi [0]{\begingroup \@sanitize@url \@Doi }%
\providecommand \@Doi  [1]{\endgroup\@@startlink{\doibase#1}\@@Doi}%
\providecommand \@@Doi [1]{#1\@@endlink}%
\providecommand \selectlanguage [0]{\@gobble}%
\providecommand \bibinfo  [0]{\@secondoftwo}%
\providecommand \bibfield  [0]{\@secondoftwo}%
\providecommand \translation [1]{[#1]}%
\providecommand \BibitemOpen [0]{}%
\providecommand \bibitemStop [0]{}%
\providecommand \bibitemNoStop [0]{.\EOS\space}%
\providecommand \EOS [0]{\spacefactor3000\relax}%
\providecommand \BibitemShut  [1]{\csname bibitem#1\endcsname}%
\bibitem [{\citenamefont {Boltzmann}(1872)}]{Boltzmann72}%
  \BibitemOpen
  \bibfield  {author} {\bibinfo {author} {\bibfnamefont {L.}~\bibnamefont
  {Boltzmann}},\ }\href@noop {} {\bibfield  {journal} {\bibinfo  {journal}
  {Wien. Ber.},\ }\textbf {\bibinfo {volume} {66}},\ \bibinfo {pages} {275}
  (\bibinfo {year} {1872})}\BibitemShut {NoStop}%
\bibitem [{\citenamefont {Cramer}\ \emph {et~al.}(2008)\citenamefont {Cramer},
  \citenamefont {Dawson}, \citenamefont {Eisert},\ and\ \citenamefont
  {Osborne}}]{Cramer_etal08}%
  \BibitemOpen
  \bibfield  {author} {\bibinfo {author} {\bibfnamefont {M.}~\bibnamefont
  {Cramer}}, \bibinfo {author} {\bibfnamefont {C.~M.}\ \bibnamefont {Dawson}},
  \bibinfo {author} {\bibfnamefont {J.}~\bibnamefont {Eisert}}, \ and\ \bibinfo
  {author} {\bibfnamefont {T.~J.}\ \bibnamefont {Osborne}},\ }\href@noop {}
  {\bibfield  {journal} {\bibinfo  {journal} {Phys. Rev. Lett.},\ }\textbf
  {\bibinfo {volume} {100}},\ \bibinfo {pages} {030602} (\bibinfo {year}
  {2008})}\BibitemShut {NoStop}%
\bibitem [{\citenamefont {Polkovnikov}\ \emph {et~al.}()\citenamefont
  {Polkovnikov}, \citenamefont {Sengupta}, \citenamefont {Silva},\ and\
  \citenamefont {Vengalattore}}]{Polkovnikov_etal}%
  \BibitemOpen
  \bibfield  {author} {\bibinfo {author} {\bibfnamefont {A.}~\bibnamefont
  {Polkovnikov}}, \bibinfo {author} {\bibfnamefont {K.}~\bibnamefont
  {Sengupta}}, \bibinfo {author} {\bibfnamefont {A.}~\bibnamefont {Silva}}, \
  and\ \bibinfo {author} {\bibfnamefont {M.}~\bibnamefont {Vengalattore}},\
  }\href@noop {} {}\bibinfo {note} {{a}rXiv:1007.5331}\BibitemShut {NoStop}%
\bibitem [{\citenamefont {Kinoshita}\ \emph {et~al.}(2006)\citenamefont
  {Kinoshita}, \citenamefont {Wenger},\ and\ \citenamefont
  {Weiss}}]{Kinoshita_etal06}%
  \BibitemOpen
  \bibfield  {author} {\bibinfo {author} {\bibfnamefont {T.}~\bibnamefont
  {Kinoshita}}, \bibinfo {author} {\bibfnamefont {T.}~\bibnamefont {Wenger}}, \
  and\ \bibinfo {author} {\bibfnamefont {D.~S.}\ \bibnamefont {Weiss}},\
  }\href@noop {} {\bibfield  {journal} {\bibinfo  {journal} {Nature (London)},\
  }\textbf {\bibinfo {volume} {440}},\ \bibinfo {pages} {900} (\bibinfo {year}
  {2006})}\BibitemShut {NoStop}%
\bibitem [{\citenamefont {Rigol}(2009)}]{Rigol09}%
  \BibitemOpen
  \bibfield  {author} {\bibinfo {author} {\bibfnamefont {M.}~\bibnamefont
  {Rigol}},\ }\href@noop {} {\bibfield  {journal} {\bibinfo  {journal} {Phys.
  Rev. Lett.},\ }\textbf {\bibinfo {volume} {103}},\ \bibinfo {pages} {100403}
  (\bibinfo {year} {2009})}\BibitemShut {NoStop}%
\bibitem [{\citenamefont {Rela{\~{n}}o}(2010)}]{Relano10}%
  \BibitemOpen
  \bibfield  {author} {\bibinfo {author} {\bibfnamefont {A.}~\bibnamefont
  {Rela{\~{n}}o}},\ }\href@noop {} {\bibfield  {journal} {\bibinfo  {journal}
  {J. Stat. Mech.},\ }\textbf {\bibinfo {volume} {2010}},\ \bibinfo {pages}
  {P07016} (\bibinfo {year} {2010})}\BibitemShut {NoStop}%
\bibitem [{\citenamefont {Reimann}(2008)}]{Reimann08}%
  \BibitemOpen
  \bibfield  {author} {\bibinfo {author} {\bibfnamefont {P.}~\bibnamefont
  {Reimann}},\ }\Doi {10.1103/PhysRevLett.101.190403} {\bibfield  {journal}
  {\bibinfo  {journal} {Phys. Rev. Lett.},\ }\textbf {\bibinfo {volume}
  {101}},\ \bibinfo {pages} {190403} (\bibinfo {year} {2008})}\BibitemShut
  {NoStop}%
\bibitem [{\citenamefont {Linden}\ \emph {et~al.}(2009)\citenamefont {Linden},
  \citenamefont {Popescu}, \citenamefont {Short},\ and\ \citenamefont
  {Winter}}]{Linden_etal09}%
  \BibitemOpen
  \bibfield  {author} {\bibinfo {author} {\bibfnamefont {N.}~\bibnamefont
  {Linden}}, \bibinfo {author} {\bibfnamefont {S.}~\bibnamefont {Popescu}},
  \bibinfo {author} {\bibfnamefont {A.~J.}\ \bibnamefont {Short}}, \ and\
  \bibinfo {author} {\bibfnamefont {A.}~\bibnamefont {Winter}},\ }\Doi
  {10.1103/PhysRevE.79.061103} {\bibfield  {journal} {\bibinfo  {journal}
  {Phys. Rev. E},\ }\textbf {\bibinfo {volume} {79}},\ \bibinfo {pages}
  {061103} (\bibinfo {year} {2009})}\BibitemShut {NoStop}%
\bibitem [{\citenamefont {Goldstein}\ \emph {et~al.}(2010)\citenamefont
  {Goldstein}, \citenamefont {Lebowitz}, \citenamefont {Mastrodonato},
  \citenamefont {Tumulka},\ and\ \citenamefont {Zangh\`\i}}]{Goldstein_etal10}%
  \BibitemOpen
  \bibfield  {author} {\bibinfo {author} {\bibfnamefont {S.}~\bibnamefont
  {Goldstein}}, \bibinfo {author} {\bibfnamefont {J.~L.}\ \bibnamefont
  {Lebowitz}}, \bibinfo {author} {\bibfnamefont {C.}~\bibnamefont
  {Mastrodonato}}, \bibinfo {author} {\bibfnamefont {R.}~\bibnamefont
  {Tumulka}}, \ and\ \bibinfo {author} {\bibfnamefont {N.}~\bibnamefont
  {Zangh\`\i}},\ }\Doi {10.1103/PhysRevE.81.011109} {\bibfield  {journal}
  {\bibinfo  {journal} {Phys. Rev. E},\ }\textbf {\bibinfo {volume} {81}},\
  \bibinfo {pages} {011109} (\bibinfo {year} {2010})}\BibitemShut {NoStop}%
\bibitem [{\citenamefont {Fermi}\ \emph {et~al.}(1955)\citenamefont {Fermi},
  \citenamefont {Pasta},\ and\ \citenamefont {Ulam}}]{FPU55}%
  \BibitemOpen
  \bibfield  {author} {\bibinfo {author} {\bibfnamefont {E.}~\bibnamefont
  {Fermi}}, \bibinfo {author} {\bibfnamefont {J.}~\bibnamefont {Pasta}}, \ and\
  \bibinfo {author} {\bibfnamefont {S.}~\bibnamefont {Ulam}},\ }\href@noop {}
  {\emph {\bibinfo {title} {Studies of non linear problems}}},\ \bibinfo {type}
  {Tech. Rep.}\ \bibinfo {number} {LA-1940}\ (\bibinfo  {institution} {Los
  Alamos},\ \bibinfo {year} {1955})\BibitemShut {NoStop}%
\bibitem [{\citenamefont {Campa}\ \emph {et~al.}(2009)\citenamefont {Campa},
  \citenamefont {Dauxois},\ and\ \citenamefont {Ruffo}}]{CamDauxRuf09}%
  \BibitemOpen
  \bibfield  {author} {\bibinfo {author} {\bibfnamefont {A.}~\bibnamefont
  {Campa}}, \bibinfo {author} {\bibfnamefont {T.}~\bibnamefont {Dauxois}}, \
  and\ \bibinfo {author} {\bibfnamefont {S.}~\bibnamefont {Ruffo}},\
  }\href@noop {} {\bibfield  {journal} {\bibinfo  {journal} {Phys. Rep.},\
  }\textbf {\bibinfo {volume} {480}},\ \bibinfo {pages} {57} (\bibinfo {year}
  {2009})}\BibitemShut {NoStop}%
\bibitem [{\citenamefont {Antoni}\ and\ \citenamefont {Ruffo}(1995)}]{AnRu95}%
  \BibitemOpen
  \bibfield  {author} {\bibinfo {author} {\bibfnamefont {M.}~\bibnamefont
  {Antoni}}\ and\ \bibinfo {author} {\bibfnamefont {S.}~\bibnamefont {Ruffo}},\
  }\href@noop {} {\bibfield  {journal} {\bibinfo  {journal} {Phys. Rev. E},\
  }\textbf {\bibinfo {volume} {52}},\ \bibinfo {pages} {2361} (\bibinfo {year}
  {1995})}\BibitemShut {NoStop}%
\bibitem [{\citenamefont {Joyce}\ and\ \citenamefont
  {Worrakitpoonpon}(2010)}]{JoyceWorrakitpoonpon10}%
  \BibitemOpen
  \bibfield  {author} {\bibinfo {author} {\bibfnamefont {M.}~\bibnamefont
  {Joyce}}\ and\ \bibinfo {author} {\bibfnamefont {T.}~\bibnamefont
  {Worrakitpoonpon}},\ }\href@noop {} {\bibfield  {journal} {\bibinfo
  {journal} {J. Stat. Mech.},\ }\textbf {\bibinfo {volume} {2010}},\ \bibinfo
  {pages} {P10012} (\bibinfo {year} {2010})}\BibitemShut {NoStop}%
\bibitem [{\citenamefont {Gabrielli}\ \emph {et~al.}(2010)\citenamefont
  {Gabrielli}, \citenamefont {Joyce},\ and\ \citenamefont
  {Marcos}}]{Gabrielli_etal10}%
  \BibitemOpen
  \bibfield  {author} {\bibinfo {author} {\bibfnamefont {A.}~\bibnamefont
  {Gabrielli}}, \bibinfo {author} {\bibfnamefont {M.}~\bibnamefont {Joyce}}, \
  and\ \bibinfo {author} {\bibfnamefont {B.}~\bibnamefont {Marcos}},\ }\Doi
  {10.1103/PhysRevLett.105.210602} {\bibfield  {journal} {\bibinfo  {journal}
  {Phys. Rev. Lett.},\ }\textbf {\bibinfo {volume} {105}},\ \bibinfo {pages}
  {210602} (\bibinfo {year} {2010})}\BibitemShut {NoStop}%
\bibitem [{\citenamefont {Kastner}(2010)}]{Kastner10}%
  \BibitemOpen
  \bibfield  {author} {\bibinfo {author} {\bibfnamefont {M.}~\bibnamefont
  {Kastner}},\ }\href@noop {} {\bibfield  {journal} {\bibinfo  {journal} {Phys.
  Rev. Lett.},\ }\textbf {\bibinfo {volume} {104}},\ \bibinfo {pages} {240403}
  (\bibinfo {year} {2010})}\BibitemShut {NoStop}%
\bibitem [{\citenamefont {Micheli}\ \emph {et~al.}(2006)\citenamefont
  {Micheli}, \citenamefont {Brennen},\ and\ \citenamefont
  {Zoller}}]{Micheli_etal06}%
  \BibitemOpen
  \bibfield  {author} {\bibinfo {author} {\bibfnamefont {A.}~\bibnamefont
  {Micheli}}, \bibinfo {author} {\bibfnamefont {G.~K.}\ \bibnamefont
  {Brennen}}, \ and\ \bibinfo {author} {\bibfnamefont {P.}~\bibnamefont
  {Zoller}},\ }\href@noop {} {\bibfield  {journal} {\bibinfo  {journal} {Nature
  Phys.},\ }\textbf {\bibinfo {volume} {2}},\ \bibinfo {pages} {341} (\bibinfo
  {year} {2006})}\BibitemShut {NoStop}%
\bibitem [{\citenamefont {O'Dell}\ \emph {et~al.}(2000)\citenamefont {O'Dell},
  \citenamefont {Giovanazzi}, \citenamefont {Kurizki},\ and\ \citenamefont
  {Akulin}}]{ODell_etal00}%
  \BibitemOpen
  \bibfield  {author} {\bibinfo {author} {\bibfnamefont {D.}~\bibnamefont
  {O'Dell}}, \bibinfo {author} {\bibfnamefont {S.}~\bibnamefont {Giovanazzi}},
  \bibinfo {author} {\bibfnamefont {G.}~\bibnamefont {Kurizki}}, \ and\
  \bibinfo {author} {\bibfnamefont {V.~M.}\ \bibnamefont {Akulin}},\
  }\href@noop {} {\bibfield  {journal} {\bibinfo  {journal} {Phys. Rev.
  Lett.},\ }\textbf {\bibinfo {volume} {84}},\ \bibinfo {pages} {5687}
  (\bibinfo {year} {2000})}\BibitemShut {NoStop}%
\bibitem [{\citenamefont {Dyson}(1969)}]{Dyson69a}%
  \BibitemOpen
  \bibfield  {author} {\bibinfo {author} {\bibfnamefont {F.~J.}\ \bibnamefont
  {Dyson}},\ }\href@noop {} {\bibfield  {journal} {\bibinfo  {journal} {Commun.
  Math. Phys.},\ }\textbf {\bibinfo {volume} {12}},\ \bibinfo {pages} {91}
  (\bibinfo {year} {1969})}\BibitemShut {NoStop}%
\bibitem [{\citenamefont {Emch}(1966)}]{Emch66}%
  \BibitemOpen
  \bibfield  {author} {\bibinfo {author} {\bibfnamefont {G.~G.}\ \bibnamefont
  {Emch}},\ }\href@noop {} {\bibfield  {journal} {\bibinfo  {journal} {J. Math.
  Phys. (N.Y.)},\ }\textbf {\bibinfo {volume} {7}},\ \bibinfo {pages} {1198}
  (\bibinfo {year} {1966})}\BibitemShut {NoStop}%
\bibitem [{\citenamefont {Radin}(1970)}]{Radin70}%
  \BibitemOpen
  \bibfield  {author} {\bibinfo {author} {\bibfnamefont {C.}~\bibnamefont
  {Radin}},\ }\href@noop {} {\bibfield  {journal} {\bibinfo  {journal} {J.
  Math. Phys. (N.Y.)},\ }\textbf {\bibinfo {volume} {11}},\ \bibinfo {pages}
  {2945} (\bibinfo {year} {1970})}\BibitemShut {NoStop}%
\bibitem [{\citenamefont {Wreszinski}(2010)}]{Wreszinski10}%
  \BibitemOpen
  \bibfield  {author} {\bibinfo {author} {\bibfnamefont {W.~F.}\ \bibnamefont
  {Wreszinski}},\ }\href@noop {} {\bibfield  {journal} {\bibinfo  {journal} {J.
  Stat. Phys.},\ }\textbf {\bibinfo {volume} {138}},\ \bibinfo {pages} {567}
  (\bibinfo {year} {2010})}\BibitemShut {NoStop}%
\bibitem [{\citenamefont {Lowe}\ and\ \citenamefont
  {Norberg}(1957)}]{LoweNorberg57}%
  \BibitemOpen
  \bibfield  {author} {\bibinfo {author} {\bibfnamefont {I.~J.}\ \bibnamefont
  {Lowe}}\ and\ \bibinfo {author} {\bibfnamefont {R.~E.}\ \bibnamefont
  {Norberg}},\ }\href@noop {} {\bibfield  {journal} {\bibinfo  {journal} {Phys.
  Rev.},\ }\textbf {\bibinfo {volume} {107}},\ \bibinfo {pages} {46} (\bibinfo
  {year} {1957})}\BibitemShut {NoStop}%
\bibitem [{\citenamefont {de~Buyl}\ \emph {et~al.}()\citenamefont {de~Buyl},
  \citenamefont {Mukamel},\ and\ \citenamefont {Ruffo}}]{BuylMukamelRuffo}%
  \BibitemOpen
  \bibfield  {author} {\bibinfo {author} {\bibfnamefont {P.}~\bibnamefont
  {de~Buyl}}, \bibinfo {author} {\bibfnamefont {D.}~\bibnamefont {Mukamel}}, \
  and\ \bibinfo {author} {\bibfnamefont {S.}~\bibnamefont {Ruffo}},\
  }\href@noop {} {}\bibinfo {note} {{a}rXiv:1012.2594}\BibitemShut {NoStop}%
\end{thebibliography}%

\end{document}